\documentclass[aps, twocolumn, amsmath,amssymb,showpacs,prx,superscriptaddress]{revtex4-2}

\usepackage{mathtools}
\usepackage{graphicx,xspace}
\usepackage{dcolumn}
\usepackage{bm}
\usepackage[dvipsnames]{xcolor}
\usepackage{hyperref}
\hypersetup{
   colorlinks=true,
   linkcolor=blue,
   filecolor=magenta,      
   urlcolor=blue,
   citecolor=blue,
}
\usepackage{float}
\usepackage{amsmath}
\usepackage{gensymb}
\usepackage{chngcntr}
\usepackage[toc,page]{appendix}
\usepackage[symbol]{footmisc}

\usepackage{chemformula}
\usepackage{siunitx}

\usepackage{pstricks}
\usepackage{dutchcal}
\usepackage{psfrag}

\usepackage[normalem]{ulem}
\usepackage{appendix}

\newcommand{\printfnsymbol}[1]{%
  \textsuperscript{\@fnsymbol{#1}}%
}
\makeatother
\usepackage{upgreek}
\usepackage{subcaption}
\usepackage[font=small,labelfont=bf,
   justification=Justified,
   format=plain]{caption}
\begin{document}

\title{Second-order nonlinear disordered photonic media}

\author{Rabisankar Samanta }

\thanks{These authors contributed equally to the work}

\thanks{Present address: Department of Ophthalmology \& Tech4Health Institute, NYU Grossman School of Medicine, NY, USA}
\affiliation{Nano-Optics and Mesoscopic Optics Laboratory, Tata Institute of Fundamental Research, 1 Homi Bhabha Road, Mumbai, 400 005, India}

\author{Romolo Savo }
\thanks{These authors contributed equally to the work}
\email{romolo.savo@cref.it}
\affiliation{Centro Ricerche Enrico Fermi (CREF), Via Panisperna 89a, 00184 Rome, Italy}

\author{Claudio Conti}
\affiliation{Centro Ricerche Enrico Fermi (CREF), Via Panisperna 89a, 00184 Rome, Italy}
\affiliation{Department of Physics, University of Sapienza, Piazzale Aldo Moro 5, 00185 Rome, Italy}
\author{Sushil Mujumdar}
\email{mujumdar@tifr.res.in}
\affiliation{Nano-Optics and Mesoscopic Optics Laboratory, Tata Institute of Fundamental Research, 1 Homi Bhabha Road, Mumbai, 400 005, India}


\begin{abstract}

The field of complex photonics has garnered significant interest due to its rich physics and myriad applications spanning physics, engineering, biology, and medicine. However, a substantial portion of research focuses primarily on the linear medium. Over the years, optical nonlinearity, particularly the second order denoted as $\chi^{(2)}$, has been harnessed for diverse applications such as frequency conversions, three-wave mixing, material characterizations, and bio-imaging. When $\chi^{(2)}$-nonlinearity combines with the disorder, a new realm of physics emerges, which in the last 30 years has witnessed substantial progress in fundamental studies and futuristic applications. This review aims to explore fundamental concepts concerning $\chi^{(2)}$-nonlinear disordered media, chart the field's evolution, highlight current interests within the research community, and conclude with a future perspective.
\end{abstract}

\maketitle
\section{Introduction}\label{intro}  
Disordered photonic media have gained significant prominence in optical science due to the many opportunities they offer for both fundamental and applied studies~\cite{wiersma2013,gigan2017,lopez2018, carminati2021,Cao2022harnessdisorder, vynck2023light}. Complex media characterized by random optical disorder, such as white paint, engineered powder aggregates, and biological tissues, have served as testbed platforms for exploring various aspects of fundamental physics. These range from the elusive Anderson localization~\cite{wiersma1997,skipetrov2016red} to anomalous diffusion processes~\cite{barthelemy2008levy} and replica symmetry breaking~\cite{ghofraniha2015}.
 The mature development of optical wavefront shaping techniques~\cite{gigan2017, cao2022} has endowed disordered photonic media with unprecedented functionalities, including light focusing~\cite{vellekoop2010}, multiplexing~\cite{Wu_Guillon_2024}, and information processing~\cite{gigan2022natphys}. 
With the advancement of disorder engineering, these media are recognized as highly versatile platforms for light management. Their sustainable, scalable, and low-cost fabrication is an additional value that makes them an attractive choice for a wide range of practical implementations~\cite{Cao2022harnessdisorder,redding2013spectro, nocentini2024}.
Furthermore, theoretical methods adapted from mesoscopic physics have provided insight into the statistical significance of optical interactions within these systems~\cite{gigan2017, vynck2023light}.

The exploration of photonic disorder in the nonlinear optics regime has progressed at a comparatively slower pace. Experimental challenges often hinder investigations, and theoretical models tend to become complex and cumbersome. However, the past three decades have witnessed substantial advancements in fundamental studies that combine disorder and nonlinearity. In particular, recent advances in disorder-based parametric down-conversion~\cite{Ru2017}, Mie-resonance enhancement of random quasi-phase matching~\cite{savo2020}, and innovative applications in machine learning and artificial neural networks~\cite{moon2023, wang2024} have reignited interest in nonlinear disordered media, tremendously widening the realm of possibilities offered by these platforms.

In general, when an optical medium responds nonlinearly to an input field E, its polarization P can be expanded in higher-order terms, as expressed in Eq.~\ref{eq:nonlinear_orders} 
\begin{equation}
\textrm{P}=\epsilon_0(\chi^{(1)}\textrm{E}+\chi^{(2)}\textrm{E}^2+\chi^{(3)}\textrm{E}^3+\dots). 
\label{eq:nonlinear_orders}
\end{equation}
Here, we specifically focus on the class of media characterized by disorder in second-order nonlinearity $\chi^{(2)}$. 
A prototypical example of such a system is a disordered assembly of noncentrosymmetric single-crystal domains with random sizes, shapes, and orientations (Fig.~\ref{fig:phase_diagram}). Despite the presence of disorder, these media allow $\chi^{(2)} \neq 0$ due to local breaking of the inversion symmetry at the molecular level within the domains.
The literature describes a variety of cases that involve different materials and scenarios. 
Although several reviews \cite{cao2022, gigan2022natphys, berlotti2022, lib2022} and books \cite{ISHIMARU-1997,carminati2021} have extensively covered linear scattering media, there is currently no comprehensive review dedicated to second-order nonlinear disordered media.
Few reviews have touched upon disordered and engineered photonic assembly for electro-optics applications \cite{vogler2022}, but they have not thoroughly addressed the overview of nonlinear disordered media.
For example, disorder may pertain only to the spatial distribution of $\chi^{(2)}$, as in transparent ferroelectric polycrystalline media, or to both $\chi^{(1)}$ and $\chi^{(2)}$, as in multiple scattering media.
In this review, our aim is to encompass all feasible configurations of second-order nonlinear disordered media and unify the different cases under a common framework, emphasizing their shared physical foundation - namely, random quasi-phase matching.
The nonlinear phenomena of interest include those typical of three-wave mixing, such as second-harmonic generation (SHG), sum- and difference-frequency generation (SFG, DFG), cascaded harmonic generation, and supercontinuum generation. 
Topics such as the speckle of second-harmonic light, diffusion, and localization of nonlinear light, wavefront shaping for optimized SHG, and recent advancements for futuristic applications, particularly in optical computing, will be addressed.
We will illustrate that the interplay between disorder and nonlinearity facilitates the emergence of hybrid and novel nonlinear optical processes. We will also discuss the challenges and significance of implementing parametric nonlinear processes in second-order nonlinear disordered photonic media, which could pave the way for the simultaneous generation and manipulation of quantum photons via disorder.
We conclude with a comprehensive discussion on the future prospects of this emerging field and the novel ideas that could influence its trajectory.

\section{From order to disorder}
Since the pioneering of nonlinear optics, bulk high-quality nonlinear crystals with a noncentrosymmetric structure have been the conventional platform for three-wave mixing. 
Because of chromatic dispersion, the efficiency and the mixing bandwidth in these media are bounded by stringent phase-matching conditions. 
The use of birefringent crystals has been the first strategy to achieve phase matching: By tuning the pump polarization angle and the crystal temperature, a pair of ordinary and extraordinary refractive indices allows propagating the linear and nonlinear optical waves with the same phase velocity, avoiding the destructive interference of the nonlinear waves.

In addition, structural engineering of the nonlinear properties of the medium has been developed as a more elaborate strategy to tailor phase-matching conditions.
Periodic poling (PP) is the simplest type of engineering of second-order nonlinearity. In PP, the ferroelectric orientation of the nonlinear crystal is regularly reversed over domains of one coherent length by applying an intense electric field over the region of interest. This results in quasi-totally constructive interference of the nonlinear waves generated from the single domains, known as quasi-phase matching (QPM)~\cite{fejer1992QPM}. Similarly to birefringent phase matching, QPM in large-thickness crystals provides very high nonlinear conversion efficiency ($\approx 50\%$)~\cite{Miller1997} and quadratic scaling of the nonlinear signal with the number of domains (Fig.~\ref{fig:rqpm}). The downsides are cumbersome fabrication, high costs, and narrow conversion bandwidth. In fact, the poling design is strictly dependent on the material and the wavelength of operation.
PP allows QPM to be reached only for a narrow cone of k-vectors around a specific single direction, i.e., it creates one-dimensional nonlinear systems.  The extensions of PP to two and even to three spatial dimensions have been a sought-after goal for many years. In practice, it corresponds to the building multi-dimensional nonlinear photonic crystals~\cite{Berger1998, Zhang2021_NPC2D3D}. Due to the many technical challenges involved, the realization of these structures has been fully achieved only recently~\cite{Bahabad2008, wei2018_NPC3D, xu2018_NPC3D}.

Relaxed phase-matching conditions and largely improved performances in broadband nonlinear mixing can be reached by increasing the complexity of the nonlinear crystal structure~\cite{suchowski2010}.
Randomly arranging crystalline $\chi^{(2)}$ nonlinear domains is an extreme version of engineering complex nonlinearity and provides most versatile performance, although at the expense of the conversion efficiency.  Additional advantages are related to the fabrication and the economic convenience. In fact, randomness and disorder spontaneously occur in low-quality materials fabricated with limited control methods, as in the natural growth of crystals and self-assembly.
%

\begin{figure}[t]
  \centering
 \includegraphics[width=\linewidth]{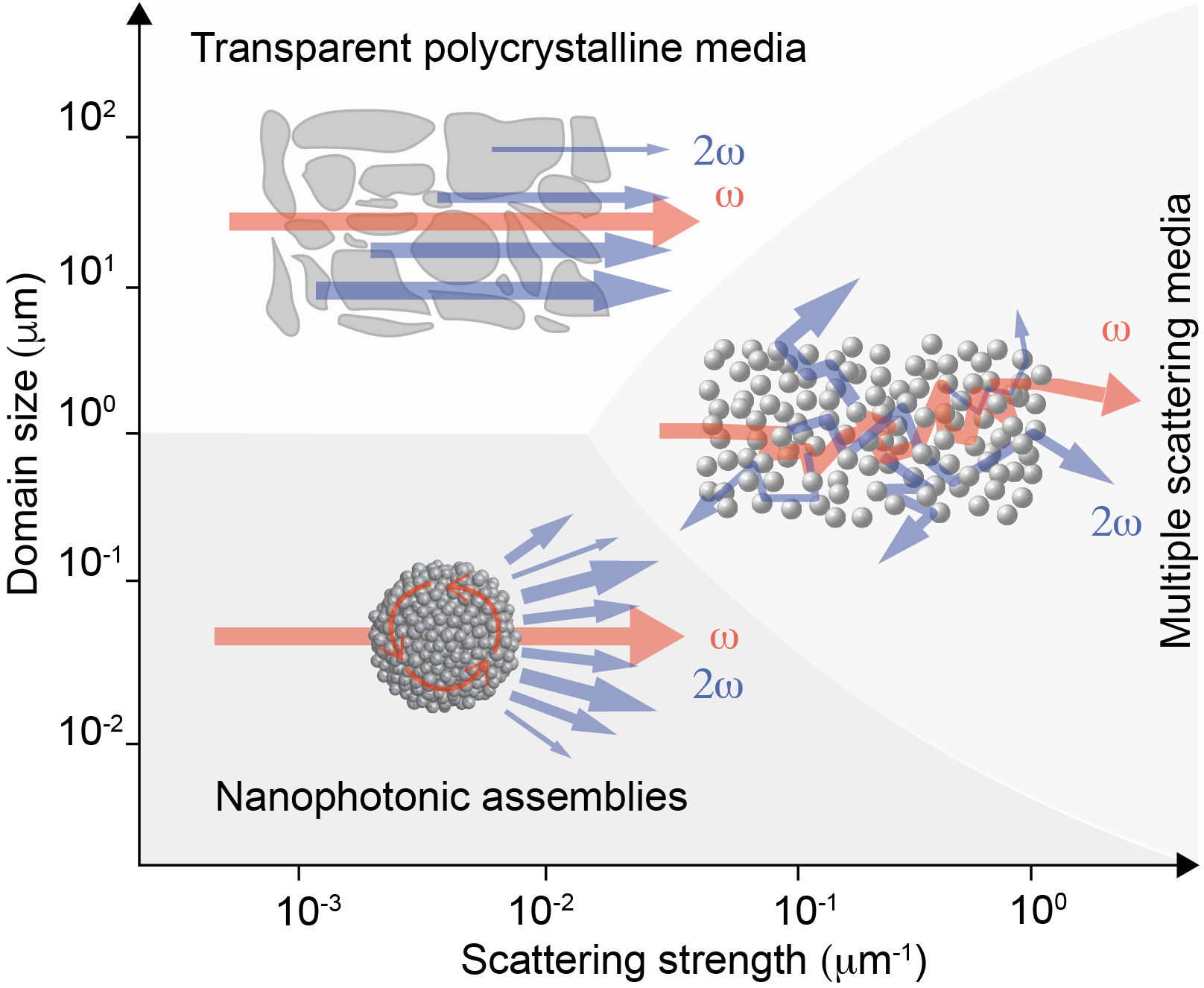}
   \caption{\textbf{Second-order nonlinear disordered media in a phase-diagram representation.} Media are classified by considering the average size of the nonlinear domain (y-axis) and the scattering strength of the disorder (x-axis). The scattering strength is calculated as the inverse of the transport mean free path at visible wavelengths.}
   \label{fig:phase_diagram}
\end{figure}
\par
A variety of second-order nonlinear disordered media can be identified when considering the average size of the nonlinear domains and the light scattering strength of the disordered structure. This type of classification is represented in Fig.~\ref{fig:phase_diagram} as a phase-diagram.
Three groups of second-order nonlinear disordered media have been investigated more significantly: (a) transparent polycrystals, having very large nonlinear domains ($10^1-10^2~\mu$m) compared to wavelengths and negligible scattering at the domain-domain interface. Since light beams can propagate freely through the disorder, they appear transparent and clear; (b) multiple-scattering media, having domain sizes of the order of the wavelength ($10^{-1} - 10^{0}~\mu$m ), are typically made up of high-index nanocrystals embedded in a low-index hosting medium (e.g. air, polymer). Under coherent light illumination, multiple scattering of light over the index inhomogeneities creates a complex interference pattern (speckle), while under broadband incoherent light illumination these media appear opaque and whitish; (c) nanophotonic assemblies are composed of sub-wavelength domains ($10^{-2}~\mu$m) creating a supra-structure having a size comparable to the wavelength. Light interacts with single domains on a smaller length scale and with the entire assembled structure on a larger length scale. 
In many practical situations, it is difficult to select arbitrary points in the diagram of Fig.~\ref{fig:phase_diagram}, as the domain size and the scattering strength are mutually dependent. 
For instance, domain sizes comparable to the light wavelength along with an index contrast at the domain surface significantly increase the scattering strength. Materials and fabrication methods also limit the variability of the two properties.

\section{ Transparent polycrystalline nonlinear media}
\begin{figure}[t]
  \centering
 \includegraphics[]{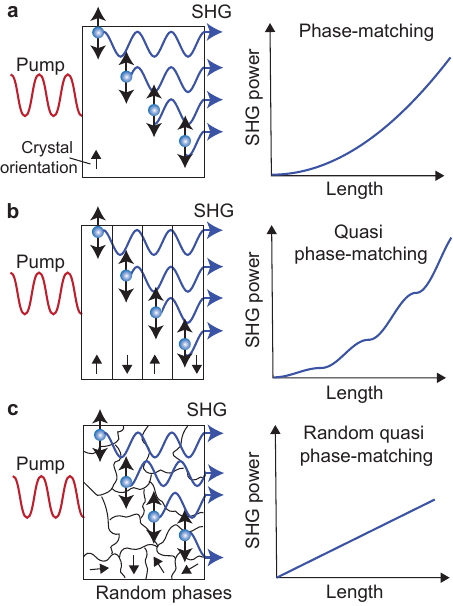}
   \caption{\textbf{Random quasi-phase matching compared to traditional types of phase-matching}. \textbf{a}, Phase-matching in birefrigent crystals, the SHG waves are generated from single electron oscillations perfectly in phase. They interfere constructively, resulting in a quadratic scaling of the SHG power with the propagation distance. \textbf{b}, Quasi-phase matching, the crystal polarization (indicated by the single arrows) is locally flipped to keep SHG waves in phase and avoid destructive interference. This results in an overall quadratic scaling of the SHG power with the propagation distance. \textbf{c}, Random quasi-phase matching, the polarization of the crystal domain is randomly oriented, SHG waves have random phase and amplitude interfering in a way that is neither destructive nor constructive. Only the intensities of the single SHG waves add up, such that the SHG power scales linearly with the propagation distance. Figure inspired by ~\cite{skipetrov2004RQPM}, adapted from~\cite{morandi2023thesis} with permission. }
   \label{fig:rqpm}
\end{figure}
\begin{figure}[t]
  \centering
  \includegraphics[width=0.95\linewidth]{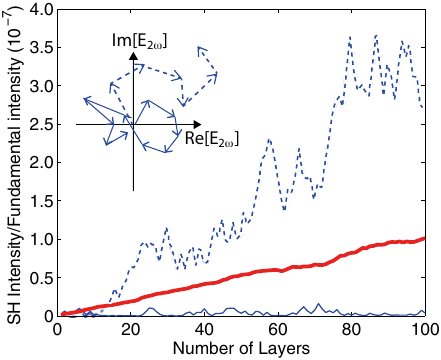}
   \caption{\textbf{Averaging process and scaling in random quasi-phase matching.} SHG efficiency calculated from the one-dimensional scalar model developed in~\cite{vidal2006}. Layers correspond to nonlinear $\chi^{(2)}$ domains. The global interference of the SHG from one configuration of disorder corresponds to a random walk in the SH-field plane (depicted in the inset). Dashed and solid blue lines report the SHG from two of these random walks. Each nonlinear layer defines a step of the random walk. Single random walk trajectories may result in very different SHG efficiencies. Only the disorder-averaged curve is linearly correlated with the number of domains (red line). The average has to be calculated over the intensity of the fields. Figure adapted from \cite{vidal2006} with permission.}
   \label{fig:rqpm_averaging}
\end{figure}

Several papers explored quasi-phase matching from the perspective of random stochastic processes. Second-order nonlinear phenomena such as SHG, SFG, and DFG were investigated in media composed of many non-centrosymmetric single-crystal domains, having random sizes, shapes, and orientations.
Along with theoretical works framing the formalisms for describing three-wave mixing in nonlinear disordered media, experiments focused on ferroelectric disordered polycrystals. These media naturally present the desired $\chi^{(2)}$ disorder at visible (VIS) and near-infrared (NIR) wavelengths, having negligible absorption (optical transparency) and negligible scattering on the domains (clear appearance).
Morozov et al. theoretically demonstrated that disordered polycrystalline samples supported 'stochastic' quasi-phase matching \cite{Morozov2001,morozov2004}.
Baudrier-Raybaut et al. reported the first experimental demonstration and modeling of 'random' quasi-phase matching in polycrystalline samples of zinc selenide (ZnSe) \cite{baudrier2004}. 
Despite differences in the approach and derivation of the results, the physical mechanism identified by the two groups is analogous: as a result of disorder, nonlinear light is generated from each domain with random amplitude and phase, the purely interference terms average out, and the total nonlinear signal is the sum of the intensities generated from each domain. The distinctive features of RQPM are depicted in Fig. \ref{fig:rqpm}: the linear (monotonic) scaling of the nonlinear optical power with the number of domains distinguishes the process from the quadratic scaling of phase matching and quasi-phase matching. 
The peculiarities of random quasi-phase matching (RQPM) have also been discussed in detail by Skipetrov S. E. in Ref.~\cite{skipetrov2004RQPM}. 
The experiment of Baudrier-Raybaut et al. observed the characteristic linear scaling of RQPM by measuring the DFG power for increasing thicknesses of ZnSe thin films, over distances much larger than the coherence length, without geometric constraints~\cite{baudrier2004}. Their result clearly showed that RQPM offers the opportunity to exploit the high $\chi^{(2)}$ components of a non-phasematchable nonlinear material, as in the case of ZnSe.
Vidal and Martorell developed a one-dimensional scalar model for RQPM and explained why the efficiency of SHG grows linearly with the thickness of the sample, or more generally, with the number of domains. They highlighted the statistical nature of the process, equivalent to a random walk in the complex plane of the SHG field. As shown in Fig. \ref{fig:rqpm_averaging}, the SHG of a single disorder configuration is not correlated with the number of domains, while the linear dependence emerges after a sufficient averaging over the disorder. They also unveiled that RQPM is expected for any size of the nonlinear domains. 
Fischer R. et al. investigated naturally grown crystals of strontium barium niobate (SBN) having a two-dimensional random distribution of the crystalline domain. They observed flat broadband ($\approx$ 50~nm) tunability of SHG using ultrafast pulses around 830~nm~\cite{fischer2006}. 
The broadband frequency conversion is a unique feature provided by $\chi^{(2)}$-disorder that arises from the absence of a resonant optimization mechanism in RQPM. Later on, Molina P. et al. reported ultrabroadband ($\approx$500~nm) SHG and THG tunability on similar SBN samples~\cite{ molina2008NLphotglass}.
In general, zinc-blende semiconductors in their disordered polycrystalline form have been extensively studied as suitable candidates for implementing RQPM, in particular when illuminated by mid-infrared (MIR) laser pulses.  
High harmonic generation and octave-spanning MIR supercontinuum were observed in polycrystalline zinc selenide (ZnSe) and zinc sulfide (ZnS)~\cite{Suminas2017,kupfer2018,Werner2019, Suminas2019}
Ultra-broadband MIR source extending from 2.7 to 20 $\mu$m was also reported from intra-pulse difference frequency generation~\cite{zhang2019RQPM}.
Ru et al. demonstrated an optical parametric oscillator (OPO) based on RQPM in polycrystalline ZnSe~\cite{Ru2017}. The OPO was pumped by femtosecond laser pulses at 2.35~$\mu$m and produced an ultra-broadband spectrum spanning 3-7.5~$\mu$m, with a pump threshold of 90~mW when using optimized samples. The OPO oscillation was achieved by illuminating the sample in correspondence to a SHG hotspot, which provided conversion efficiencies 2-3 times higher than those of the average RQPM process. This remarkable achievement represents the first OPO that utilizes a $\chi^{(2)}$ nonlinear disordered medium and the first OPO based on ZnSe.

RQPM in transparent polycrystals has also been considered as a means to generate frequencies at the limit of the photonic spectral region. Trabs P. et al.~\cite{Trabs2016} reported the generation of tunable coherent radiation in the vacuum ultraviolet (UV) down to 121~nm using RQPM in strontium tetraborate. Wang S. et al.~\cite{wang2022rectif} modeled optical rectification based on RQPM in ZnSe ceramics to convert ultrashort pulses to broadband terahertz (THz) waves. 

\subsection{Random quasi-phase matching: modeling}
\begin{figure}[t]
  \centering
  \includegraphics[width=\linewidth]{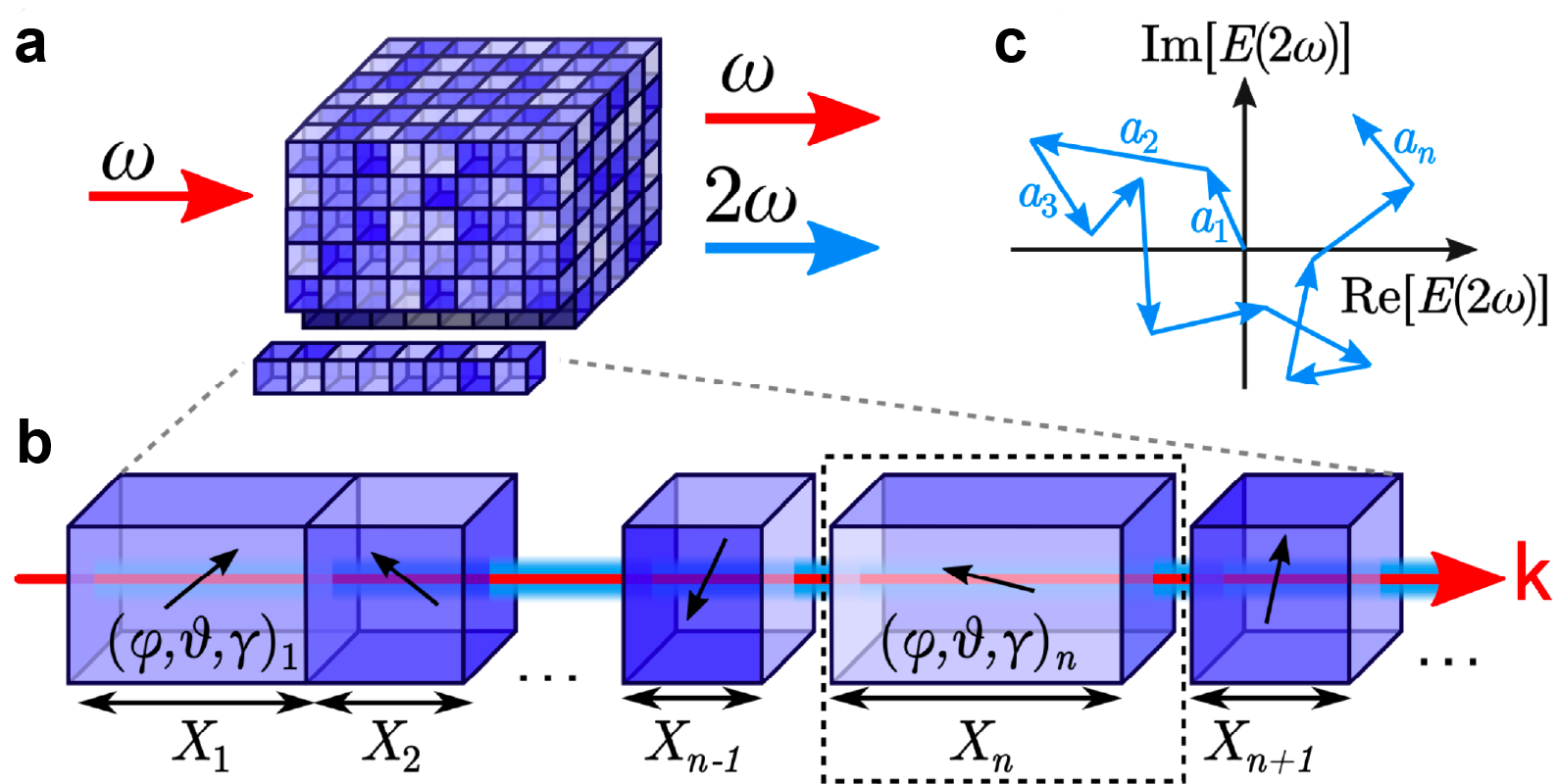}
   \caption{\textbf{Modeling scheme of random quasi-phase matching for second-harmonic generation.} \textbf{a}, Sketch of a three-dimensional disordered assembly of second-order nonlinear domains, defining multiple parallel one-dimensional sticks. \textbf{b}, Each stick contains nonlinear domains of varying size $X_n$ and crystal orientation $(\phi, \theta, \gamma)_n$, indicated by the arrow and the variation in color. \textbf{c}, Phasor ($a_n$) representation of the interference between the second-harmonic waves generated by the domains within the stick.  Each stick corresponds to a single random walk in the SHG complex plane, in which the step length is the amplitude of the second-harmonic field. \textbf{a}-\textbf{c} adapted from~\cite{muller2021} with permission.}
   \label{fig:rqpm_modeling}
\end{figure}

Random quasi-phase matching models have been developed to describe three-wave mixing processes in transparent disordered polycrystals. Therefore, both light propagation and nonlinear generation are considered forward-directed, without interaction orthogonal to the pump direction. In other words, the three-dimensional disordered nonlinear medium is seen as a cuboid structure consisting of many quadratic ($\chi^{(2)}\neq 0$) crystalline domains. Typical free parameters are the average size and degree of polydispersity of the nonlinear domains. A representation of the model is shown in Fig. \ref{fig:rqpm_modeling}a.
 The three-dimensional problem is reduced to a one-dimensional problem by breaking down the cuboid into multiple parallel one-dimensional sticks. The nonlinear generation is calculated separately along each stick, as shown in Fig. \ref{fig:rqpm_modeling}b for the case of SHG. 
 The field at the end of a stick is the sum of the contributions from each domain. The generated amplitudes and phases depend on the material and the disorder.  This interference process is equivalent to a random walk in the SHG complex plane, as represented schematically in Fig. \ref{fig:rqpm_modeling}c. 
 The random walk picture allows one to immediately identify the origin of the linear power scaling with the sample thickness observed experimentally. The SHG power of a polycrystalline sample corresponds to the mean square displacement of this random walk, which scales linearly with the number of steps, that is, the number of domains in a stick. Averaging over many sticks corresponds to averaging over disorder.
In the following, we review the models developed to calculate the nonlinear field generated from the single domains. 
\subsubsection{Scalar model }
The first approach to modeling RQPM relied on the scalar approximation of both the optical field and the $\chi^{(2)}$ susceptibility of the domains.  
Referring to the case of SHG, the generated field $E(2\omega,X_n)$ at the end of the $n^{th}$ domain of length $X_n$ is given by
\begin{equation}
     E(2\omega,X_n) \propto P^{(2)}(2\omega, X_n) (\frac{e^{i\Delta k X_n}-1}{\Delta k})e^{ik_3X_n},
  \label{eq:RQPM_singledomain_scalar}
\end{equation}
with $P^{(2)}(2\omega, X_n )=\epsilon_0\chi^{(2)}(2\omega, X_n)E^2(\omega)$ the second-order polarization. In the previous expressions, $\chi^{(2)}(2\omega, X_n )$ is the second-order nonlinear susceptibility of the $n_{th}$ domain, $E(\omega)$ is the pump field, $\Delta k=k(2\omega)-2k(\omega)$ the phase mismatch, $k_3=k(2\omega)$ is the wave vector of the second-harmonic. The expression in brackets in Eq.~(\ref{eq:RQPM_singledomain_scalar}) is the classical SHG growth term for a single crystal~\cite{boyd2008}. 

Using the scalar approximation, Baudrier-Raybaut et al.~\cite{baudrier2004} derived analytic expressions for disorder-averaged observables, namely the DFG intensity and the effective number of domains participating in the DFG process. In that work, it was not considered necessary to provide an explicit expression for the scalar $\chi^{(2)}(2\omega,X_n)$ appearing in eq.~(\ref{eq:RQPM_singledomain_scalar}), which requires the dependence on the domain orientation. They used an effective nonlinear susceptibility  $\chi^{(2)}_\textrm{eff}$ corresponding to the average of the scalar $\chi^{(2)}(2\omega,X_n)$ over all possible domain orientations.
Morozov et al.~\cite{Morozov2001, morozov2004} explicitly parameterized the second-order nonlinear susceptibility as $\chi^{(2)}(X_n)=\phi(n)\chi^{(2)}$ within a random telegraph process, that is, by using $\phi(n)=\pm 1$. This description is specific for SHG from polycrystals with needlelike, antiparallel domain structure, such as that of naturally grown SBN~\cite{kawai1998second,fischer2006}. 
Vidal and Martorell parameterized the scalar nonlinear susceptibility in the most general way, letting $\phi(n)$ have any possible value between -1 and 1 with equal probability~\cite{vidal2006}. Specifically, they used $\phi(n)=cos(\pi y_n)$ with $y_n$ uniformly selected in the interval $[0,1]$. They conducted a thorough study of the effect of domain sizes on the RQPM process. As shown in Eq.~\ref{eq:RQPM_singledomain_scalar}, the domain size distribution plays a key role in randomizing the phases of the second-harmonic waves.
Contrary to what was assumed in previous works~\cite{baudrier2004, kawai1998second}, they evidenced that the characteristic linear growth of the intensity of the SHG with the number of domains is expected for any average size of the domains, not only when the size of the domains approaches the coherence length of the material~\cite{baudrier2004}. The domain size affects the RQPM efficiency (i.e., slope of the linear intensity scaling), which is maximal when domain sizes get closer to the coherence length of the material.
Chen X. and Gaume R. provided a more explicit discussion of the scalar model developed in~\cite{vidal2006} and used it to describe SHG from non-stoichiometric grain-growth ZnSe ceramics.
In general, despite its simplicity, the scalar model captures important underlying mechanisms of RQPM and provides good qualitative agreement with the experimental results.
\subsubsection{Vectorial model for isotropic polycrystals}
More comprehensive vectorial models have been developed to take into account light polarization in the RQPM process and enable more quantitative comparisons with experiments. In this description, each component of the SHG field at the end of the single domain is given by
\begin{equation}
     E_i(2\omega,X_n) \propto P_i^{(2)}(2\omega, X_n) (\frac{e^{i\Delta k X_n}-1}{\Delta k})e^{ik_3X_n},
  \label{eq:RQPM_singledomain_vect_iso}
\end{equation}
with $P_i(2\omega, X_n) = \epsilon_0\sum_{jk} \chi^{(2)}_{ijk}(2\omega,X_n)E_j(\omega)E_k(\omega)$. The indices refer to the principal axes in a crystal coordinate system, and the second-order nonlinear susceptibility $\chi^{(2)}_{ijk}(2\omega,X_n)$ is a third-rank tensor.
Kavamori et al. investigated the case of SHG from isotropic (non birefringent)  Zinc-Blende polycrystals~\cite{kawamori2019}. 
In their study, the susceptibility variation due to the random orientation of the nonlinear domain is modeled by randomly rotated crystal coordinates $(x, y, z)$ with respect to the laboratory frame $(x', y', z')$.
The operation is implemented as the product of three rotation matrices $R=R(\theta)R(\phi)R(\psi)$, one for each Euler angle. 
The nonlinear polarization in the laboratory frame is then computed as ${\chi^{(2)}}'_{i'j'k'}=R^T_{i'i}R^T_{j'j}R^T_{k'k} \chi^{(2)}_{ijk}$. 
By implementing this approach in a Monte Carlo routine they derived the exact distribution of the nonlinear coefficients for both output SHG polarizations, relying on a realistic distribution of domain sizes and orientations measured on ZnSe polycrystalline samples.
The model was verified by direct comparison of the histograms of the measured SHG intensities with those calculated numerically.
Gu et al. extended the vectorial model to the temporal dimension to study the supercontinuum generation from ultra-short pulses in the mid-infrared in ZnSe polycrystals~\cite{ Gu2020Simulation} and the effect of RQPM on pulse coherence properties \cite{Gu2021rqpm_pulse}.
\subsubsection{Vectorial model for birefringent polycrystals}
Birefringent single crystals are the common means to achieve phase matching in three-wave mixing processes. The orientation- and polarization-dependent refractive index allows for compensating the phase lag accumulated by the mixing waves along the propagation, obtaining constructive interference of all generated nonlinear waves.
M\"uller et al. developed a vectorial model to simulate SHG in birefringent polycrystals, taking into account both the full three-dimensional rotation and the birefringence of $\chi^{(2)}$ domains ~\cite{muller2021}.
In the crystal frame of a single domain, the beams are decomposed into their components along the ordinary ($o$) and extraordinary ($e$) axes and each pump polarization combination ($oo,eo,oe,ee$) generates a second harmonic field $E^u(2\omega,X_n)$ along $u\in \{o,e\}$, which is given at the end of the $n^{th}$ single domain according to
    \begin{equation}
            E^u(2\omega, X_n) \propto \sum_{v,w} P^{u,vw}(2\omega, X_n) \bigg(\frac{e^{i\Delta k^{u,vw}X_n}-1}{i\Delta k^{u,vw}}\bigg)e^{ik^u_3X_n}.
            \label{eq:RQPM_singledomain_vect_bire}
    \end{equation}
With the phase mismatch $\Delta k^{u,vw} = k^v(\omega) + k^w(\omega) - k^u(2\omega)$, the wave vector $k^u_3$ of the second-harmonic along $u$, and $P^{u,vw} = \langle\hat{\text{\bf{e}}}^{u}, \bm{P}^{vw}\rangle$ the second-harmonic polarization along $v,w\in\{o,e\}$ projected onto the $o/e$-axis ($\hat{\text{\bf{e}}}^{u}$ unit vector along $u\in\{o,e\}$), where $P_i^{vw}(2\omega, X_n) = \epsilon_0\sum_{jk}\chi^{(2)}_{ijk}(2\omega, X_n)E_j^vE_k^w$. 
The model calculates the generation and propagation of the waves from domain-to-domain, considering the phases of the ordinary and extraordinary beam components separately. This is done  within the approximation of no scattering at the domain-to-domain boundaries.
M\"uller et al. used the model to simulate SHG from LiNbO$_3$, BaTiO$_3$ and ADP disordered polycrystals in both non-phase-matchable and phase-matchable conditions. Although the influence of birefringence is not appreciable for smaller domains, birefringence starts to play a notable role once the domains are larger than the coherence length of the material~\cite{muller2021}.
In polydispersed polycrystals with sufficiently large domains, birefringence introduces an enhancement in SHG efficiency of up to 54$\%$ compared to isotropic reference crystals. Birefringence also supports RQPM in monodispersed polycrystalline assemblies, which can outperform the SHG efficiency of a polydispersed assembly in the large-domain regime.

\subsubsection{Extension to resonant polycrystalline structures}
\label{sec:RQPM_resonantpoly}
Savo et al.~\cite{savo2020} extended the RQPM model to the case of a disordered polycrystalline structure that sustains optical modes, as the nanophotonic nonlinear assemblies discussed in Section~\ref{sec:NanoPhotNonlinAss}.  
The model addresses the SHG in a one-dimensional disordered array of $\chi^{(2)}$ domains in the large-scale modes approximation, i.e., when the spatial features of the modes at $\omega$ and $2\omega$ evolve on a scale larger than the mean size of the domains. The SHG fields $E_i(2\omega, X_n)$ of the individual domains are first calculated as described in the previous sections. To account for the non-uniform pump distribution of the modes, a domain-dependent SHG enhancement factor $\xi_i(\omega, 2\omega)=\omega^2 F_i^2(\omega)F_i(2\omega)$ is multiplied by the SHG field, which enhances or decreases the SHG of the domain accordingly. $F_i(\omega)$ and $F_i(2\omega)$ are the field enhancements of the pump and of the SHG, respectively. 
In this case, the randomly quasi-phase matched SHG power is given by the mean squared displacement of a random walk in the SHG plane having a mode-dependent step-length distribution, in which some steps contribute more than others if the corresponding domain is in a high-enhancement region.  
For the case of spherical assembly of nonlinear nano-domains sustaining Mie modes, the expression of the total SHG power from the structure is
\begin{equation}
\label{eq:RQPM_resonant}
   P_\mathrm{SHG}(R,\omega,2\omega) = \eta ({\chi^{(2)}_\mathrm{eff}\omega})^2 N(R) I(\omega)^2 \cdot \overline{\xi^2}(R,\omega,2\omega) 
\end{equation}
where $\eta$ is a proportionality factor that accounts for the RQPM efficiency for a given mean size and polydispersity of the domains, $\chi^{(2)}_\mathrm{eff}$ is the effective second-order susceptibility of the material, $N(R)$ is the number of domains within a microsphere, $I(\omega)$ is the pump intensity, and $\overline{\xi^2}(R,\omega,2\omega)$ is the volume-averaged squared field-enhancement of the Mie modes.
 
\section{Multiple-scattering nonlinear disordered media}

Essentially, we have so far dealt with transparent media, with weak scattering that arises mostly due to crystal impurities. There are no distinct boundaries between the various scattering centers, so naturally, index preservation leads to maintenance of the direction of propagation. The physics of transport in this case is determined primarily by the domain size. Now, we move into the realm of disorder where distinct, clear scattering centers can be identified. Essentially, a scattering particle is involved, which is characterized by the scattering cross-section, which depends on the input wavelength and particle size. A conglomerate of such scattering particles forms the disordered system characterized by the transport mean free path, the average distance over which the direction of propagation is randomized. The physics of transport depends on the relative magnitude of the transport mean free path and the sample size.  Generally,  early exiting light requires samples thicker than 8 - 10 transport mean free paths to reach complete diffusion \cite{kop1997,svensson2013,carminati2021}. Naturally, for a nonlinear sample with two wavelengths co-existing, the transport can be different for the two components in the same sample. This feature is observed for all mesoscopic parameters of interest, such as, for example, speckle statistics and speckle correlations. Although speckle formation depends on the phases of the scattered light, we note that the transport per se is not dominated by the phase. That situation leads to exotic interference effects such as weak localization and Anderson localization. The latter requires scattering conditions that, to our knowledge, have not been met in nonlinear disorder hitherto. A more robust characterization of the diffusive strength is provided by the exponentially-decaying tail of the transmission in the temporal regime. Notably, apart from the strength of disorder, the strength of nonlinearity is also a significant criterion of transport. Weak nonlinearity is assumed to exist when the intensity of the fundamental light does not deplete with propagation. Strong nonlinearity, on the other hand, modulates the intensity distribution of the fundamental light in the sample. 
The forthcoming text discusses certain aspects of light diffusion in nonlinear disorder. Specifically, we elaborate on nonlinear scattering and spectroscopy, diffusion and localization, speckle statistics and correlations. We also add a subsection discussing nonlinear characterization of such samples, since the development of bulk crystals with macrosizes is severely challenging and expensive for certain materials. 

\subsection{Nonlinear light scattering and spectroscopy}

The combination of nonlinear effects and light scattering can be traced back to 1965 when a group of scientists studied scattered SHG light by focusing a ruby laser inside water \cite{terhune1965}. This study later laid the groundwork for the field of nonlinear light scattering (NLS) and spectroscopy. Depending on the size of the scatterers, NLS can be categorized into different groups. If the scatterer is less than 10 nm in size, it is identified as Hyper-Rayleigh Scattering (HRS). A detailed review of NLS and spectroscopy can be found in Ref. \cite{roke2012}.

NLS and spectroscopy have primarily focused on liquid samples containing scatterers such as droplets, emulsions, and particles. NLS is a powerful method for identifying the location and size of micro-structures buried within a medium \cite{beer2009}. However, NLS is based mostly on scattering from single particles or small clusters. For example, researchers have been able to distinguish between scattering from single particles and clusters by analyzing angle-dependent nonlinear light, specifically the sum-frequency generated light intensity \cite{dadap2009}. However, the combination of nonlinear scattering and simultaneous photon transport in a $\chi^{(2)}$-nonlinear disordered medium exhibits different phenomena, which are thoroughly discussed in the following sections.

\subsection{Diffusion and localization}

 In a medium with a random distribution of refractive indices, photons experience multiple scattering events. As the strength of the disorder increases, photon transport undergoes a transition from ballistic to diffusive and even localized behavior \cite{segev2013}. The scenario becomes particularly intriguing when nonlinearity becomes a factor. Disordered media with $\chi^{(2)}$-nonlinearity govern the transport for two types of photons: the fundamental one and the second harmonic one generated inside the fundamental beam via random quasi-phase-matching. The simultaneous generation of second harmonic light and its subsequent transport has garnered significant interest. In the late '80s, the coherence backscattering (CBS) effect of light in linear complex media, considered a signature of weak localization \cite{albada1985,wolf1985, maret1987}, created a buzz in the photonics community. Consequently, researchers sought the same phenomenon in SHG. Agranovich and Kravtsov were the first to theoretically investigate weak localization in a second-order nonlinear disordered medium for there-wave mixing \cite{agranovich1988}. In their theoretical study, they chose a doped semiconductor as the nonlinear disordered sample, in which scattering of the fundamental photons was considered negligible. However, SHG photons were thought to experience multiple scattering events in the sample. They observed an enhanced peak of SHG light in the backward direction, where the peak appeared mainly due to the photons generated from the surface layer (within transport mean free path, $\ell_t$). However, the magnitude of the CBS peak was found to be much smaller than that of linear disordered media. They also observed that, for sample thickness $L>>\ell_t$, a diffuse background of the SHG light could hinder the visibility of the weak localization in steady state, which would be only visible on a very small time scale. Subsequently, Kravtsov et al. extended their theoretical approach in the case of four-wave mixing \cite{kravtsov1990}. In the experimental domain, Yoo et al. tried to see the CBS of SHG in a fine powder of KDP microcrystals \cite{yoo1989}. Unfortunately, they could not succeed, which they attributed to the specific strength of the disorder in the nonlinear medium. In the following years, Kravtsov et al. \cite{kravtsov1991} tried to address the shortcomings in the experiment performed by Yoo et al. \cite{yoo1989}. This time, they considered strong scattering for both harmonics, unlike their earlier study \cite{agranovich1988}, where only SHG scattering was considered. They proposed that the absence of CBS peak of SHG photons in the experiment was mainly due to two reasons: (a) $\langle \chi^{(2)}\rangle=0$, and (b) the scatterers' size was much smaller than the transport mean free path at SHG wavelength i.e. $R<<\ell_{2\omega}$. In subsequent years, researchers successfully observed the CBS of SHG light experimentally in surface plasmonic systems \cite{simon1992} and silver films \cite{wang1993}. After several years, Makeev et al. provided a theoretical overview of photon diffusion in a colloidal suspension of spherical nonlinear particles, with the assumption that the pump did not deplete with the propagation in the medium \cite{makeev2003}. Later, Faez et al. investigated experimentally the spatial diffusion of SHG photons in GaP powders and presented diffusion equations for SHG light \cite{faez2009}.  The self-trapping of optical beams with random quasi-phase matching was reported in Ref. \cite{Conti2010}. A purely nonlinear localization mechanism was proposed in Ref. \cite{Folli2013}. In a recent study, researchers reported cavity-enhanced SHG in strongly disordered media composed of LiNbO$_3$ nanoparticles \cite{qiao2019}. 

\begin{figure*}[!t]
   \centering
   \includegraphics[width = \linewidth]{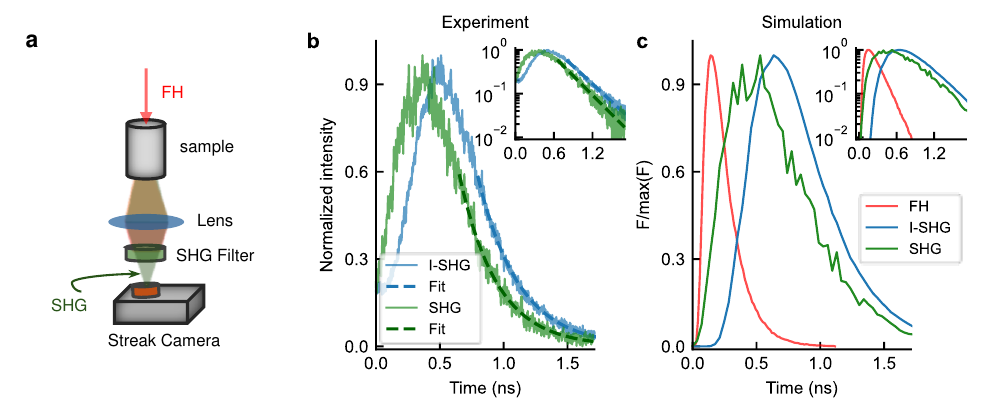}
   \caption{\textbf{Temporal diffusion of light in a $\chi^{(2)}-$nonlinear disordered medium}. \textbf{a,} Schematic of the experiment, a pulsed laser was made incident normally onto the sample with cylindrical geometry, and the transmitted SHG light (after filtering) was analyzed by a streak camera. \textbf{b,} Experimental results. The diffusion of SHG photons (solid green line) compared with the diffusion of incident SHG (I-SHG, solid blue line) photons coupled to the medium from outside. The tails of diffusion profiles are fit (dashed lines) to exponential decay profiles. Inset: Diffusion profiles plotted in log scale showing linear tail. \textbf{c,} Numerically simulated temporal diffusion profiles of FH (solid red line), I-SHG (solid blue line), and SHG (solid green line). Inset: Numerically simulated temporal diffusion profiles in log-scale. Figures are recreated from Ref. \cite{samanta2023}  with permission, American Physical Society.}
   \label{fig:nonlin_diff}
\end{figure*}

        
        

Samanta et al. investigated spatial and temporal photon diffusion in a second-order nonlinear disordered medium of finite-size cylindrical geometry \cite{samanta2023}. The sample was a pellet packed with KDP microcrystals. Their observations revealed that the spatial intensity profile of the SHG peaked deeper inside the medium compared to the fundamental beam. To compare the diffusion of linear photons at the same wavelength of SHG, they coupled an externally generated SHG beam (termed as incident SHG or I-SHG) onto the sample. Their study revealed that although SHG and I-SHG are of the same wavelength, their diffusion profiles peak at different depths. I-SHG peaks early because its wavelength is shorter than that of FH. In contrast, SHG peaks deepest inside the sample, as sSHG photons are continuously generated inside the disordered medium from the FH photons.  The most significant part of their work was the study of temporal diffusion for the first time (see the schematic of the experiment in Fig.~\ref{fig:nonlin_diff}a). As seen in their experimental results (Fig.~\ref{fig:nonlin_diff}b), the diffusion of SHG peaked early compared to I-SHG in the temporal domain. The temporal diffusion of the FH could not be measured owing to the spectral constraint of the streak camera. In theory, they deduced the diffusion of SHG photons in the long-time limit as 
\begin{equation}
F_{\textsc{SHG}}(t) \propto 
\begin{cases}
  \frac{e^{-\Gamma(2\omega)t}-e^{-2\Gamma(\omega)t}}{2\Gamma(\omega)- \Gamma(2\omega)} & \text{if $\Gamma(2\omega) \ne 2\Gamma(\omega)$} \\
  te^{-\Gamma(2\omega)t} & \text{otherwise}
\end{cases} 
\end{equation}
where $\Gamma(\omega)$ and  $\Gamma(2\omega)$ are the decay rates for the FH and SHG, respectively. They numerically solved the temporal diffusion profiles for all three types of photons and observed a similar (see Fig.~\ref{fig:nonlin_diff}c) trend as the experiment. The diffusion of FH peaks early in time, as the longer wavelength experiences less scattering than the SHG or I-SHG. More importantly, as the birth places of the SHG photons are distributed inside the medium and also depend on the diffusion of FH, they arrive earlier at the transmission side than the I-SHG photons.

\subsection{Speckle of nonlinear light: statistics and correlation}
\begin{figure*}[htbp]
   \centering
   \includegraphics[width = \linewidth]{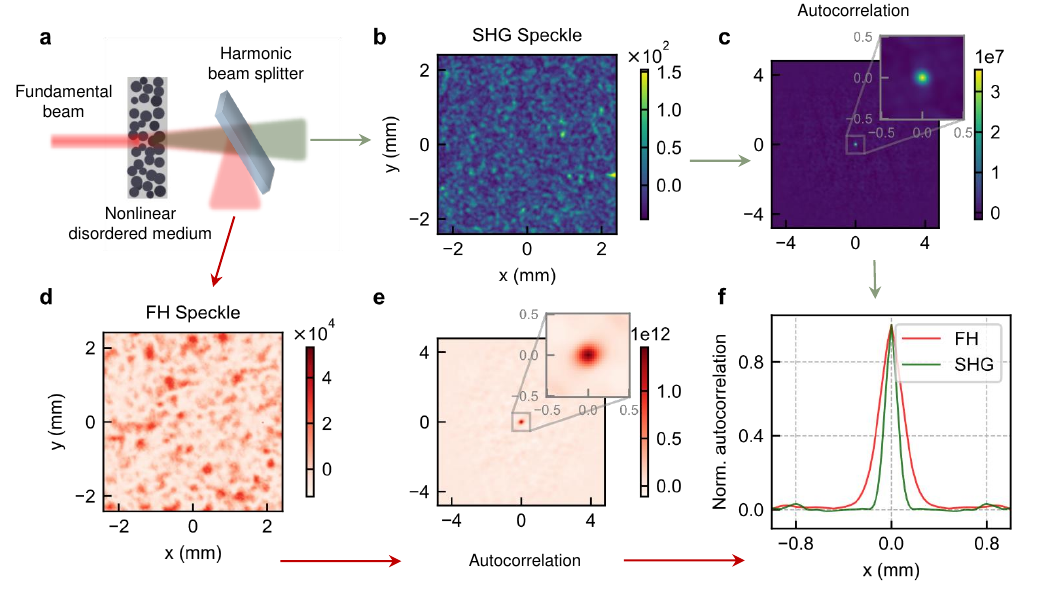}
   \caption{\textbf{Speckle patterns in a second-order nonlinear disordered medium.} \textbf{a}, The nonlinear disordered medium generates the speckle patterns of fundamental and second harmonic light, which are separated by a harmonic beam splitter. \textbf{b}, SHG Speckle \textbf{d}, FH speckle. \textbf{c} and \textbf{e} are the 2D autocorrelation map of \textbf{b} and \textbf{d} respectively. Zoomed-in images highlight the autocorrelation peak. \textbf{f}, Normalized autocorrelation peak along the X-axis for FH (red) and SHG (green) light.}
   \label{fig:FH_SH_speckle}
\end{figure*}
\begin{figure*}[hbtp]
   \centering
   \includegraphics[width=\linewidth]{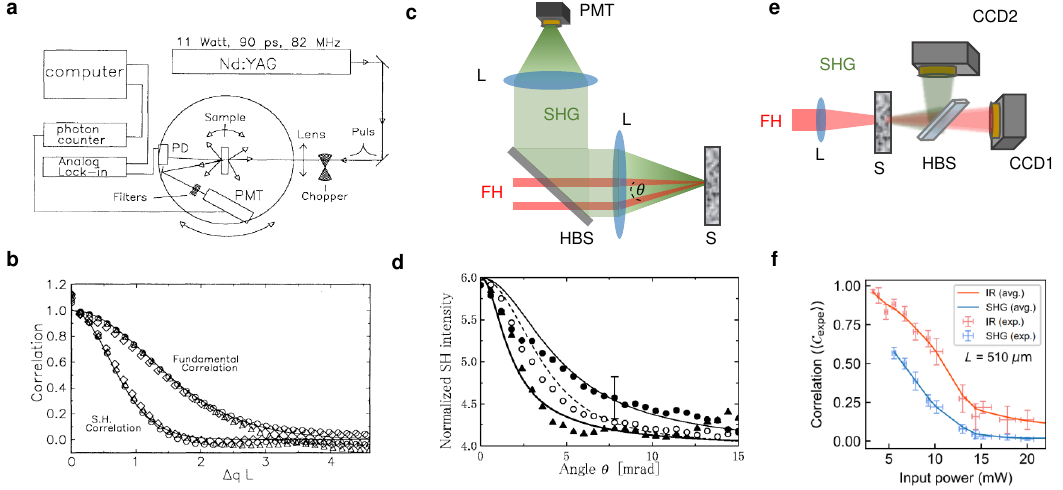}
   \caption{\textbf{Speckle correlation in second-order nonlinear disordered media.} Notations: PD, Photodiode; PMT, Photomultiplier tube; L, Lens; HBS, Harmonic Beam Splitter; S, Nonlinear scattering sample;  CCD1, Charge-Coupled Device with Si detector;  CCD2, Charge-Coupled Device with InGaAs detector; \textbf{a,} Schematic of the experimental setup for measuring speckle correlations of FH and SHG in transmission and reflection as a function of sample rotation. \textbf{b,} Speckle correlation of FH and SHG light in transmission for three different sample thicknesses. Scatter points are the experimental data, while solid lines are the theoretical curves. \textbf{c,} Schematic of experiment where the variation of SHG intensity was measured as a function of angle ($\theta$) between two fundamental beams. \textbf{d,} Normalized SHG intensity plotted as a function of $\theta$. Solid and open markers are for reflection and transmission measurements, respectively.  Circles: sample thickness 97~$\mu$m, triangles: sample thickness 270~$\mu$m. Solid lines are theoretical curves. \textbf{e,} Schematic of experiment where the speckle correlation was measured as a function of input power of the fundamental beam.  \textbf{f,} Correlation in consecutive speckles for both harmonic drops as the input power was increased. Markers are the experimental data, and the solid lines are the average line obtained after fitting a spline to the experimental data. (\textbf{a, b}), (\textbf{c, d}) and (\textbf{e, f}) are recreated and adapted with permission from Ref. \cite{boer1993}, \cite{ito2004} and \cite{samanta2022} respectively, American Physical Society.}
   \label{fig:speckle_corr}
\end{figure*}

When light travels through a disordered medium composed of linear scatterers, speckle with random dark and bright spots appears at transmission or reflection, due to the interference effect among the coherence fields \cite{goodman2007}. Unlike linear media, a $\chi^{(2)}$-nonlinear disordered medium creates speckles for both harmonics when ultrashort laser pulses with high peak intensity pass through the medium (see Fig.~\ref{fig:FH_SH_speckle}a). In Fig.~\ref{fig:FH_SH_speckle}b and d, we show two example speckle patterns for SHG and FH, respectively, which are captured at the same place and same distance from the nonlinear disordered sample. The average speckle size, generally obtained from the width (FWHM or 1/e) of the autocorrelation peak \cite{goodman2007}, is depicted in Fig.~\ref{fig:FH_SH_speckle}c, e, f. As expected from the properties of a speckle, that is, the average size of the speckle grains increases linearly with wavelength \cite{goodman2007}, is also observed in this case. The average speckle grain size of the FH light (red) is almost twice that of SHG (green) (Fig.~\ref{fig:FH_SH_speckle}f). The following subsections discuss speckle correlations and statistics in FH and SHG.

 \subsubsection{Correlations in SHG speckle}

 Boer et al. experimentally measured short-range correlations in SHG light generated within nonlinear disordered media using transmission and reflection geometries \cite{boer1993}. Specifically, they evaluated the correlations for both harmonics while rotating the sample, which comprised LiNbO$_3$ particles ranging in size from 0.1 $\mu$m to 5 $\mu$m (see the schematic of the experiment in Fig.~\ref{fig:speckle_corr}a). Additionally, they computed the short-range correlation functions and demonstrated outstanding agreement with the experimental findings (Fig.~\ref{fig:speckle_corr}b). The most notable outcome of their study was the observation that, in reflection, the correlation for the SHG light was proportional to the sample thickness. This stands in stark contrast to linear complex media, where the correlation for the fundamental light scales with the transport mean path of the medium. \par 

Ito and Tomita measured SHG speckle correlations in disordered media of LiNbO$_3$ microparticles \cite{ito2004}. They excited the samples with two fundamental beams of the same wavelength. While one beam was sent normally to the sample, the angle of incidence ($\theta$) of the other beam on the sample was continuously varied, and subsequently the total SHG intensity was measured (see the schematic of the experiment in Fig.~\ref{fig:speckle_corr}c). They observed that when the angle between two excitation beams was zero, the total SHG intensity was maximum, while it gradually decreased with the angle between the two beams (see Fig.~\ref{fig:speckle_corr}d). They showed that the maximum intensity at $\theta \sim 0$ was due to the mutual correlation between the SHG speckle patterns generated within the medium. As the angle of incidence of one beam was changed, the SHG speckle associated with it also changed.  \par

Samanta et al. studied speckle correlations for both harmonics in $\chi^{(2)}$-nonlinear disordered media while varying the intensity of the fundamental beam \cite{samanta2022}. The sample utilized in the experiment consisted of KDP microparticles, loosely packed inside the sample (see the schematic of the experiment in Fig.~\ref{fig:speckle_corr}e). The researchers observed speckle decorrelation in both the fundamental and SHG light as the intensity of the input beam increased. Specifically, they measured the correlation between the initial two consecutive speckles for both harmonics simultaneously. Additionally, regardless of the input laser power, the correlation was consistently higher in the fundamental speckle compared to the SHG speckle (Fig.~\ref{fig:speckle_corr}f). This decorrelation phenomenon was attributed to the gradual change in particle configuration, influenced by the highly intense laser pulses. To substantiate the experimental findings, a comprehensive theoretical investigation was conducted.

\subsubsection{Statistics of SHG speckle}
In a linear disordered medium, intensities of a fully developed speckle are found to follow Rayleigh statistics \cite{goodman2007}. In the past four decades, this topic has not only been explored for the fundamental aspects and even found in several applications \cite{dainty2013}. In recent years, researchers have utilized non-Rayleigh speckle statistics for specific purposes in different domains of physics \cite{dholakia2011,jendrzejewski2012,kuplicki2016}, and also were able to create customized speckles with desired intensity statistics on a specific plane \cite{bromberg2014,bender2018} or in 3D \cite{SeungYun2023}.

Unfortunately, there are a handful of instances in which the statistics of SHG speckle have been discussed. Savo et al. presented intensity statistics of SHG speckles emerging from disordered microspheres where each microsphere was composed of a large number of BaTiO$_3$ nanocrystals (more on nanophotonic assemblies discussed in Sec. \ref{sec:NanoPhotNonlinAss}) \cite{savo2020}. In their experiment, they observed Rayleigh statistics in the SHG speckles for large microspheres (diameters larger than 5~$\mu$m), where SHG speckles evolved from a large number of fully randomized emitters. Later on, Samanta et al. studied global and local speckle contrast, as well as their statistics as a function of fundamental light intensity, for both harmonics \cite{samanta2020}. They observed that while the statistics of the speckle contrast of FH obey the existing theory, the SHG speckle shows a slight deviation. In another study, reported by Nardi, Morandi, and colleagues in a recent preprint, statistics of speckle intensity and total transmission in nonlinear disordered media were discussed \cite{nardi2024}. Their research revealed how the statistics of diffuse light, particularly for SHG light, can be influenced by altering the illumination conditions of FH. For example, they demonstrated that the speckle intensity statistics exhibit a super-Rayleigh distribution when the nonlinear disordered medium is illuminated with a tightly focused beam. This deviation was more pronounced for SHG light. However, outside the focused illumination area, the statistics tend to follow a Rayleigh distribution. Similarly, the statistics of total transmission (for both harmonics) deviate from a Gaussian distribution under focused illumination.

\subsection{Estimation of nonlinear coefficients from the nonlinear powder sample}

Estimation of nonlinear coefficients ($d_{ij}=\chi_{ij}/2$) for any nonlinear material is of significant importance in the field of nonlinear optics. Typically, the SHG efficiency of a single crystal serves as the basis for calculating these coefficients. However, in numerous instances, the fabrication of a single crystal of the desired size remains a challenge. To address this issue, Kurtz and Perry introduced a technique that mitigates the limitation \cite{kurtz1968}. Their approach relied on nonlinear powder samples, which are more readily available. Through semi-quantitative measurements, they estimated $d_{ij}$ and offered an assessment of the feasibility of phase-matching. The Kurtz-Perry (KP) method has gained widespread popularity due to its ability to provide a rough estimate of $d_{ij}$ for new nonlinear materials, as evidenced by its substantial presence in Google Scholar (accumulating more than 6000 citations in 2024). However, discrepancies often arise between the $d_{ij}$ values obtained by the KP method and those obtained from measurements on single crystals, which differ by an order of magnitude. This discrepancy is probably due to the exclusion of inherent scattering effects in the KP model. Additionally, the utilization of an index-matching liquid, a requirement for the KP method, may not always be feasible for all types of nonlinear samples. In the early 90s, Kiguchi et al. demonstrated a new method to measure the efficiency of SHG in powder samples, namely second harmonic wave generated with the evanescent wave (SHEW) \cite{kiguchi1992, kiguchi1994}. This technique could measure nonlinear coefficients for both single-crystal and powder samples.

More than three decades after the renowned KP method, Aramburu et al. took the initiative to address some of the previously mentioned issues \cite{aramburu2011}. They made slight modifications to the KP method, resulting in a more reliable estimation of the $d_{ij}$ values. To mitigate the effects of scattering, they utilized a monolayer of nonlinear particles. Subsequently, Aramburu and their colleagues proposed an alternative method to determine $d_{ij}$ values with greater accuracy \cite{aramburu2014_apb, aramburu2014}. In this instance, no index-matching was necessary, and experiments were conducted on thick and dry powder nonlinear samples.

Recently, M\"uller et al. showed that the KP model can not be applied to nano-grains or grains whose sizes are smaller than the coherence length. The KP model distinguishes the phase-matchability and non-phase-compatibility of a material based on the dependence of SHG on the grain size. This relationship cannot be directly applied to nano-grains, where the dependence of SHG on the grain size is solely material-specific, and is governed by the $\chi^{(2)}$-tensor of the specific grain. Later, Chowdhury et al. realized a technique based on SHG microscopy to analyze powder samples \cite{chowdhury2016}. In contrast to earlier methods \cite{kurtz1968, aramburu2011, aramburu2014}, they employed a focused Gaussian laser beam for SHG, rather than a collimated Gaussian beam.

\section{Nanophotonic nonlinear assemblies}
\label{sec:NanoPhotNonlinAss}
\begin{figure*}[!hbtp]
   \centering
   \includegraphics[width=0.95\linewidth]{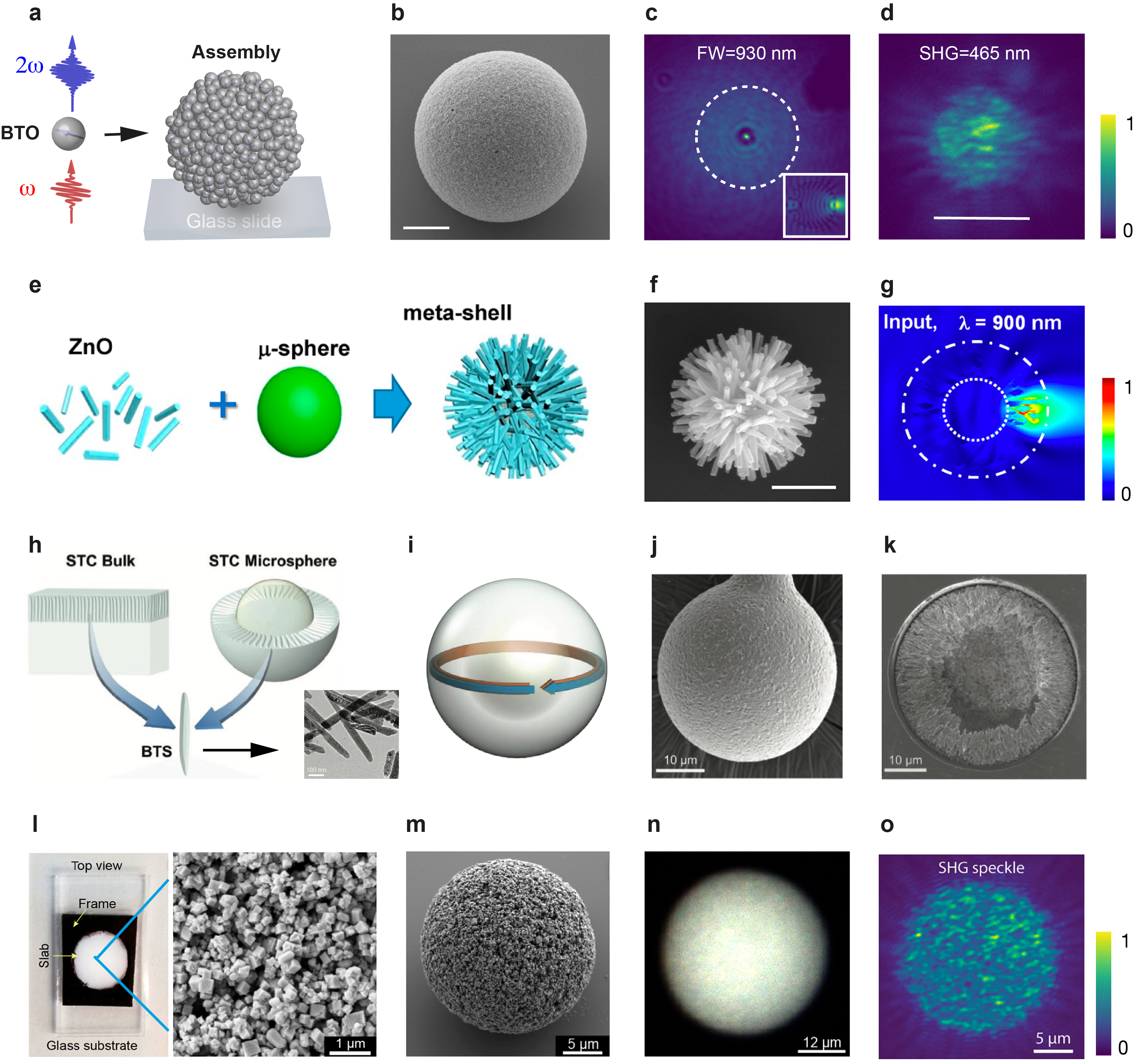}
   \caption{\textbf{ Nanophotonic nonlinear disordered systems realized by the assembly of $\chi^{(2)}$ nanocrytals into geometry-controlled suprastructures.} \textbf{a}, Schematic of the assembly of barium titanate (BTO) nano-crystals into spherical suprastructures. BTO units individually generate the second-harmonic. \textbf{b}, Scanning electron microscopy (SEM) image of an assembled  BTO micro-sphere. Scale bar 2~$\mu$m. \textbf{c}, Optical transmission image of the rear plane of a microsphere (diameter 3.2~$\mu$m ) at the fundamental wavelength (FW) 930~nm. The micro-sphere creates a bright photonic nanojet (center), enhancing the SHG. The inset shows a section of the corresponding nanojet mode (simulated). \textbf{d}, Image of the second-harmonic generation (SHG) from a BTO microsphere with a diameter of 5.1 ~$\mu$m (white line).
   \textbf{e}, Schematic of the assembly of a zinc oxide (ZnO) meta-shell supraparticle (MSP). ZnO nanorods attach to a dielectric micro-sphere to form a spherical disordered array (meta-shell); \textbf{f}, SEM image of an assembled ZnO MSP. Scale bar 1~$\mu$m. \textbf{g}  FDTD simulation, at  $\lambda$=900~nm, showing the photonic nanojet enhancing the SHG from the MSP.   
   \textbf{h}, Schematic of the surface texture crystallization (STC) glass ceramic (GC) samples. The SEM image shows the Ba$_2$TiSi$_2$O$_8$ (BTS) microcrystals forming a shell around the glass core of the microsphere. \textbf{i}, Schematic of the whispering gallery (WG) mode of the pump (orange) and of the SHG (blue) enhancing the random quasi-phase matching. \textbf{j},  SEM image of the STC GC microshere surface.\textbf{k}, SEM image of the cross section of a STC GC microsphere after corrosion.
  \textbf{l}, Photograph of a nonlinear disordered slab (variable thickness 0.5-5~$\mu$m) assembled from lithium niobate (LNO) nanocubes (SEM image in the inset). \textbf{m}, SEM image of a nonlinear disordered micro-sphere assembled with the same LNO nanocubes. \textbf{n}, Dark-field microscopy image of an LNO assembled micro-sphere. The white appearance is a clear indication of multiple scattering of light. \textbf{o}, Image of the SHG from an LNO assembled microsphere. \textbf{a}-\textbf{d} adapted from~\cite{savo2020}, \textbf{e}-\textbf{g} adapted from~\cite{Bahng_Marandi2020}. \textbf{h}-\textbf{k} adapted from~\cite{Chen2024}, \textbf{l}-\textbf{o} adapted from~\cite{morandi2022}. 
  }
   \label{fig:nanophotonic_assemblies}
\end{figure*}
In this section, we review studies related to second-order nonlinear disordered nano- and micro-structures. 
We focus on colloidal assemblies of $\chi^{(2)}$ -nonlinear nanophotonic elements in superstructures with controlled geometry.
In this class of systems, the emergence of collective geometric resonances and mode confinement enhances the nonlinear interaction~\cite{pu2010coreshell, Lin2016NLoverlap}, balancing the low conversion efficiency due to the limited amount of nonlinear material and the disorder. Remarkably, nanophotonic nonlinear disordered assemblies enable the implementation of RQPM in a miniature system. 

\subsection{All-dielectric assemblies}
Savo et al.~\cite{savo2020} realized all-dielectric three-dimensional $\chi^{(2)}$ disordered micro-spheres by bottom-up assembly of barium titanate (BaTiO$_3$) nano-crystals and explicitly demonstrated their SHG through RQPM (Fig.\ref{fig:nanophotonic_assemblies} (a-d)). The average size ($\approx$ 50~nm) and packing density ($\approx$ 55$\%$) of nanocrystals were appropriate to create an effective refractive index at the pump and the SHG wavelength ($n_\textrm{eff} \approx$ 1.55). Thanks to the optical homogeneity in the linear regime, assembled microspheres sustain high-order Mie resonances associated with \textit{photonic nanojet} mode confinement~\cite{chen2004nanojet}. A combination of broadband and resonant SHG was observed by sweeping the pump from near-infrared to mid-IR wavelengths, indicating coupling of RQPM with the geometric resonances of the micro-spheres. The nanojet mode enhances the SHG by one order of magnitude while RQPM relaxes the phase matching conditions. These BaTiO$_3$ disordered micro-spheres are as efficient as crystalline BaTiO$_3$ of the same size, despite a 71$\%$ reduction of the amount of material. The observed RQPM was modeled by a non-trivial random walk in the SHG complex plane, as described by Eq.~\ref{eq:RQPM_resonant} in Section~\ref{sec:RQPM_resonantpoly}. 
Almost simultaneously, Bahng H. J. et al.~\cite{Bahng_Marandi2020} demonstrated enhanced SHG from core-shell supraparticles fabricated by assembling zinc oxide (ZnO) nanorods into a spherical array around dielectric micro-spheres (Fig \ref{fig:nanophotonic_assemblies} (e-g)). The linear scattering properties of the assembly are determined by the radial graded effective index and the structural anisotropy of the shell. Fabrication parameters are used to engineer the Mie resonances and obtain a photonic nanojet spatially overlapping the $\chi^{(2)}$-nonlinear shell of ZnO. At 1550~nm pump wavelength, an absolute SHG conversion efficiency of $10^{-7}$ was measured. Because of the circular symmetry of the structure and the disordered arrangement of the ZnO nanorods, the SHG efficiency is independent of the input polarization direction.  
Chen J. et al.~\cite{Chen2024} observed an 80-fold enhancement of SHG by coupling RQPM with the whispering gallery modes (WGM) of a glass ceramic micro-cavity. They realized a spherical core-shell structure of a few tens of micrometers through surface crystallization of glass containing elongated ($\approx$ 1$~\mu$m) Ba$_2$TiSi$_2$O$_8$ (BTS) micro-crystals (Fig.~\ref{fig:nanophotonic_assemblies}h-k). The core, made of a transparent glass precursor, is surrounded by a disordered polycrystalline BTS layer resulting from the assembly of the BTS micro-crystals. The assembled layer is sufficiently homogeneous to support the WGM of the pump on an ultra-wide range of wavelengths, from 860~nm to 1600~nm. No significant variation of the SHG enhancement was observed over the wavelength range thanks to the flat spectral response of RQPM. A sharp SHG polarization dependence was reported and attributed to the structural anisotropy of the BTS micro-crystals, which is not sufficiently averaged by the disorder. Notably, pumping in the mid-infrared allowed them to observe cascaded sum-frequency generation in the visible range.
Bringing the size of the nonlinear crystallites close to the wavelength has dramatic effects on the optical properties of the assemblies. Morandi et al.~\cite{morandi2022} assembled LiNbO$_3$ nanocubes ($\approx$ 400~nm) to realize nonlinear disordered structures of different geometries. They produced disordered slabs (Fig.~\ref{fig:nanophotonic_assemblies}l) with a continuously variable thickness (0.3-5~$\mu$m) and disordered micro-spheres (Fig.~\ref{fig:nanophotonic_assemblies}m-o) of varying diameter (2-13~$\mu$m), showcasing simultaneously strong multiple scattering of light and SHG. They demonstrated that the RQPM is independent of the geometry of the multiple-scattering medium.   
The systems examined so far highlight template assembling~\cite{kim2008microspheres,vogel2015color, marino2024nanoassembly} of pre-synthesized nonlinear $\chi^{(2)}$ crystallites or with in situ crystallization of the nonlinear units as a powerful, facile, and scalable approach to engineer nonlinear photonic disorder at the nanoscale and microscale. Solution processing combined with nanoimprint lithography is another valuable method to fabricate nonlinear photonic structures~\cite{karvounis2020BTO, vogler2022}. Vogler-Neuling et al. have realized nonlinear woodpile photonic crystals from solution-processed disordered polycrystalline barium titanate, showing promising results for large surface area (cm$^2$) applications~\cite{VoglerNeuling2020woodpilePC, VoglerNeuling2024woodpilePC}.
\subsection{Hybrid metal-dielectric assemblies}
The assembly of hybrid metal-dielectric nanophotonic structures has allowed coupling in nonlinear photonic disorder with plasmonic resonances. Zhong et al.~\cite{zhong2020hybrid} observed sum-frequency generation with relaxed polarization dependence from disordered gold (Au) nanosponges infiltrated with ZnO nanocrystals. Inversely, Ali et al.~\cite{ali2023hybrid} reported a 32-fold enhancement of the SHG from mesoporous LiNbO$_3$ particles coated with a dispersed layer of 10~nm diameter gold nanoparticles without introducing finely tailored radiative nanoantennas to mediate photon transfer to or from the nonlinear material. 
Nanophotonic nonlinear disordered structures share several features with meta-atoms~\cite{Tonkaev2024}, sometimes without sharp boundaries between the two~\cite{Bahng_Marandi2020}. Common aspects are the use of subwavelength nanoparticles to build super-structures exhibiting effective-medium features and collective large-scale resonances. An explicit difference is that the former do not require sub-wavelength nanoparticles to be resonant, nor to be monodispersed, nor to be positioned with a specific order.

\section{Controlling scattered nonlinear light via wavefront shaping}

The field of optical wavefront shaping through scattering media began with a seminal work by Vellekoop and Mosk in 2007, in which they successfully focused light through a layer of paint using an SLM \cite{vellekoop2007}. In the last two decades, the field of complex photonics has witnessed tremendous efforts in controlling light via wavefront shaping \cite{gigan2022}, mostly through linear disordered media. With the introduction of various wavefront shaping tools such as liquid crystal spatial light modulators (LC-SLM), digital micromirror devices (DMD), and deformable mirrors (DM), researchers have not only managed to mitigate the effects of scattering but have also leveraged scattering media for various applications, including optical computing \cite{gigan2022natphys}, micro-manipulation \cite{cao2022}, and quantum information processing \cite{lib2022}. In recent years, some research groups have ventured into the field of wavefront shaping in nonlinear disordered systems, showing control over nonlinear light and enhanced computing capabilities compared to the linear counterpart. In the following subsections, we will discuss various aspects of wavefront shaping in second-order nonlinear disordered media, which has become a rich and exciting domain of research.


\begin{figure*}[t]
   \centering
   \includegraphics[width=\linewidth]{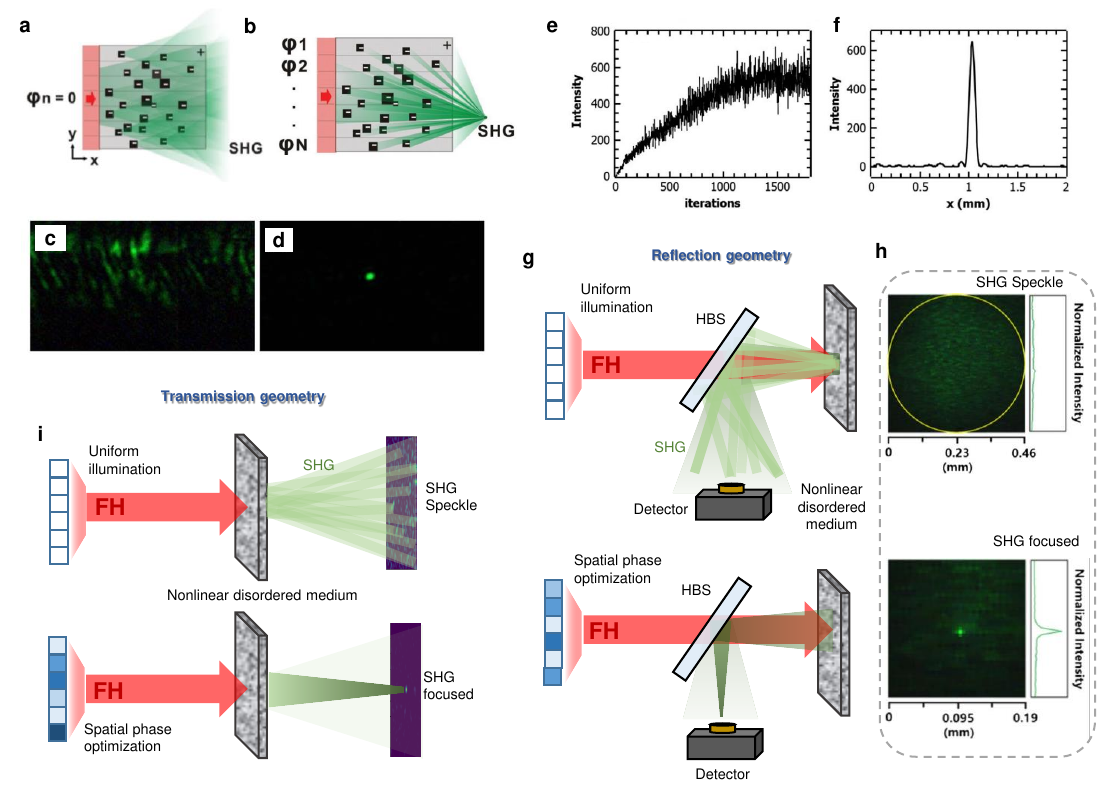}
   \caption{\textbf{Focusing of scattered SHG via feedback-based wavefront shaping.} \textbf{a,} Without optimization of the fundamental beam's phase, the random polycrystalline nonlinear crystal produces a diverged SHG beam. \textbf{b,} The phase-optimized fundamental beam makes the scattered SHG light focused at a target location. \textbf{c,} The speckle of SHG light for condition mentioned in \textbf{a}. The focused SHG beam for optimized illumination. \textbf{e,} Enhancement of the SHG intensity at the target location with iteration numbers. \textbf{f,} Intensity at the target location has been enhanced more than 600 times compared to the background intensity. 
 \textbf{g,} Schematic of the wavefront shaping to focus back-scattered SHG light emerging from a diffusive nonlinear medium. \textbf{h,} Top panel: Experimental image of an SHG speckle encircled by a yellow line. Bottom: A clear enhancement of the SHG light at the target location as confirmed by the line profiles on the right side. \textbf{i,} Schematic of the wavefront shaping to focus scattered SHG light generated from a diffusive nonlinear medium at the transmission geometry. \textbf{a, b} adapted from Ref. \cite{yao2013}, American Physical Society, \textbf{c-f} adapted from Ref. \cite{yao2012}, \textbf{h} adapted from Ref. \cite{qiao2017} with permission, Optica Publishing Group.}
   \label{fig:SH_speckle_focus}
\end{figure*}

\subsection{Enhancing and focusing scattered SHG via feedback-based optimization}

The wavefront shaping technique was first applied to transparent polycrystalline nonlinear crystals with randomly sized domains and antiparallel polarizations. Yao et al. demonstrated experimentally and theoretically that the speckle pattern of the SHG light, generated from a transparent nonlinear crystal of strontium barium titanate with domains of opposite polarizations and varying sizes, could be converted into a focused beam by optimizing the phase of the fundamental beam \cite{yao2012, yao2013}. In their theoretical framework, they calculated the SHG intensity from a single domain and then computed the total intensity by coherently summing the intensities from all the domains. Under plane-wave illumination, an SHG speckle pattern is observed, while optimizing the phase of the input fundamental wave enhances the SHG intensity in the desired direction (see Fig.~\ref{fig:SH_speckle_focus}a, b). In the experiment, they employed a partition algorithm \cite{vellekoop2008} to convert the SHG speckle pattern into a single focused spot (see Fig.~\ref{fig:SH_speckle_focus}c-f). On the other hand, a single crystal with scattering defects, also regarded as a weakly scattering medium, suffers from scattering losses. As a result, the efficiency of the second harmonic light can not reach the ideal value (without any scattering defects). Wang et al. \cite{wang2018_shg} demonstrated that optimizing the phase of the fundamental beam by a spatial light modulator (SLM) could significantly enhance the SHG efficiency ($\sim 1.15$ times during phase-matching condition ). In their experiment, a single crystal of BaMgF$_4$ was utilized as the weakly scattering nonlinear sample, and for the feedback-based optimization, they adopted a stepwise sequential algorithm \cite{vellekoop2008}.\par

In contrast, a diffusive nonlinear medium diverges the SHG light, making it difficult to collect and focus on a target location with plane wave illumination (see the schematic in Fig.~\ref{fig:SH_speckle_focus}g top panel). With a careful phase optimization of the fundamental wavefront, the scattered SHG can be focused at a desired location (see the schematic in Fig.~\ref{fig:SH_speckle_focus}g  bottom top panel). Qiao et al. \cite{qiao2017} explored a diffusive nonlinear material comprising LiNbO$_3$ nanocrystals, and achieved focused scattered light in the reflection geometry. A genetic algorithm \cite{conkey2012} was used for the feedback-based optimization (Fig.~\ref{fig:SH_speckle_focus}h). Furthermore, their study showed that the optimized fundamental beam increased the overall SHG intensity. However, until now, there has been no report of focusing the scattered SHG in transmission geometry (Fig.~\ref{fig:SH_speckle_focus}i) utilizing feedback-based optimization. The challenge here lies in the poor transmission of the diffuse SHG light. Wu et al. observed enhanced SHG in the scratched region within a thin film of BaMgF$_4$ nanocrystals. The accumulation of numerous nanoparticles in this scratched area supported extended optical path lengths, which ultimately facilitated longer interactions \cite{wu2018}.

\begin{figure*}[t]
   \centering
   \includegraphics[width=0.95\linewidth]{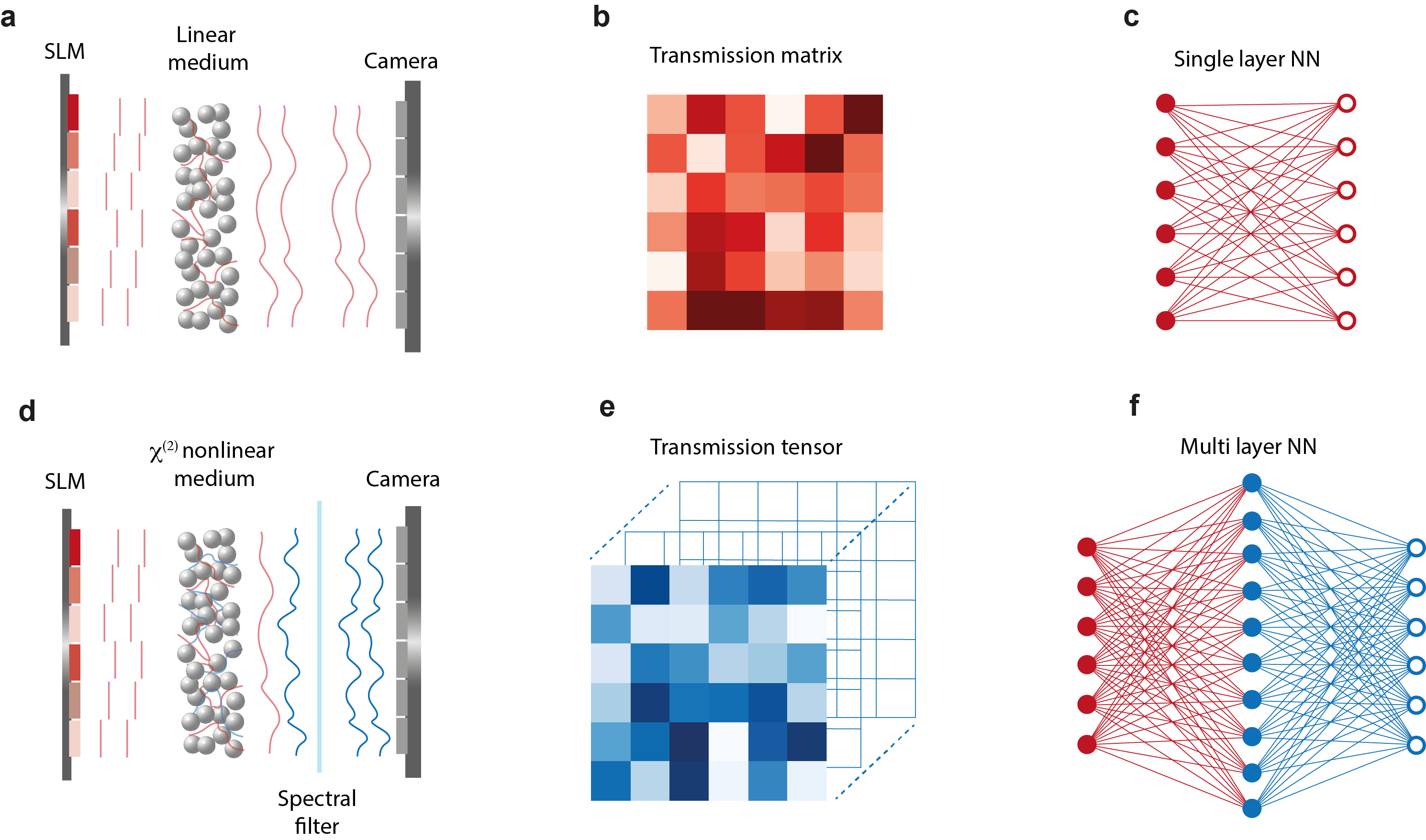}
   \caption{\textbf{Optical computing with linear and nonlinear disordered photonic media.} \textbf{a}, Schematic of the experimental configuration adopted for measuring the transmission matrix of a linear disordered medium. N pixels of the spatial light modulator (SLM) define the input vector, and N pixels of the detection camera define the output vector. The transmission matrix contains the complex number that couples the input and output. \textbf{b}, Visualization of the real part of the measured transmission matrix (N$\times$N). The color map indicates random numbers normalized to the interval between 0 (dark red) and 1 (white). \textbf{c}, Architecture of the feed-forward single-layer neural network (NN) implemented optically in \textbf{a}.  The coupling coefficients are the elements of the transmission matrix shown in \textbf{b}. The activation function is applied on the output layer of the NN by the camera detection (empty red circles). \textbf{d},  Schematic of the experimental configuration adopted for measuring the transmission tensor (i.e., transmission part of the scattering tensor) of a nonlinear ($\chi^{(2)}$ ) disordered medium. Different from the linear case, camera pixels detect the SHG, not the pump light. \textbf{e}, Visualization of the real part of the transmission tensor. The color map indicates a random number normalized to the interval between 0 (dark blues) and 1 (white). \textbf{f}, Simplified architecture of the feed-forward two-layer neural network (NN) implemented optically in \textbf{d}.  The coupling coefficients are the elements of the transmission tensor shown in \textbf{e}. The activation function is applied all-optically on the first layer of the NN layers through SHG (blue circle) and electro-optically on the output layer of the NN by the camera detection (empty blue circles). }
   \label{fig:NL_computing}
\end{figure*}

\subsection{Spectral control of broadband SHG}

The use of femtosecond pulses, which are essentially broadband in nature, enhances the efficiency of SHG. However, spectral control becomes challenging due to limitations imposed by phase-matching conditions for broadband sources. While quasi-phase-matching efficiently produces SHG for a broadband laser in highly nonlinear crystals, diffusive nonlinear media create spectral speckles in the SHG light. When broadband light traverses a linear disordered medium, it generates spatio-spectral or spatio-temporal speckles upon transmission \cite{weiner2011}. During the past decade, various methods have been employed to control broadband light spatio-spectrally or spatio-temporally through a linear disordered medium \cite{katz2011, mccabe2011, aulbach2011, mounaix2016, mounaix2016_2, mounaix2017, mounaix2018}. In most cases, the input beam was shaped using a liquid crystal-based spatial light modulator \cite{small2012}, while in a few instances, a digital micromirror device was used \cite{samanta2023_DMD}. Similarly, a diffusive nonlinear medium creates spatio-spectral speckle patterns for both harmonics. Qiao et al. took an approach to control the spectral nature of the broadband speckle of SHG light \cite{qiao2018}. They randomly selected any wavelength from the output spectra and enhanced the intensity at that wavelength by using feedback-based wavefront shaping of the fundamental beam.


\subsection{Scattering tensor}
Till now, we have discussed feedback-based wavefront optimization of fundamental light to control scattered SHG. However, this approach is inefficient and time-consuming, as for different operations, one has to iterate again to find the optimal wavefront. Estimation of a transmission matrix of a scattering medium, which directly connects the outputs to the inputs, solved the problems \cite{popoff2010}. In linear scattering media, the transmission matrix ($T_{ij}$) is expressed as $E_i^{\textrm{out}}=\sum_jT_{ij}E_j^{\textrm{in}}$. A similar kind of parameter was unavailable for the non-linear scattering medium until the significant discovery by Moon et al. \cite{moon2023}. They showed that the input-output response of a nonlinear scattering medium can be fully demonstrated by estimating a three-dimensional scattering tensor, as opposed to the two-dimensional transmission matrix in linear scattering media. 
The main analogies and differences between the scattering tensors of a $\chi^{(2)}$-nonlinear disorder medium and the transmission matrix of a linear disordered medium are represented in Fig.~\ref{fig:NL_computing}.  
To define a scattering tensor, consider a nonlinear scattering medium consisting of $n$ particles randomly placed within the sample. The numbers of input and detection channels are $N$ and $\alpha$, respectively. Then the SHG signal generated on the $\alpha$ output channel is defined as $E_{\alpha}^{\textsc{SHG}} = \sum_{j,k}^Nd_{\alpha jk}E_{j}E_{k}$, where ${d}_{\alpha {jk}}={\sum }_{l=1}^{n}{s}_{\alpha l}{\chi }_{l}{t}_{{lj}}{t}_{{lk}}$, and ${d}_{\alpha {jk}}={d}_{\alpha {kj}}$ \cite{moon2023}. $\chi_l$ are the second-order nonlinear susceptibilities. On the other side, multiple linear scattering of SHG waves from it's generation point at the particle $l$ to the $\alpha$ detection channel is defined by a linear scattering operator, termed $s_{\alpha l}$. The term ${d}_{\alpha {jk}}$ eventually forms the elements of the scattering tensor $D$. The three-dimensional scattering tensor (D) can be reshaped as a two-dimensional tensor as $D_s$, which means $E_{\alpha}^{\textsc{SHG}} = D_s \textbf{S}$, where \textbf{S} is composed of elements $S_p$ with $S_p =E_jE_k$. With the help of the scattering tensor, they were able to retrieve the input fundamental wave from the SHG output wave. To do the inverse process, they first inverted the reshaped scattering tensor ($D_s$) to get \textbf{S} and finally utilized \textbf{S} to get $E_{in}$. 


 \subsection{Optical computing}
The estimation of a 3-dimensional scattering tensor opened up a new era in the domains of information and computing.  If we compare the rank of the scattering tensor and the transmission matrix for N orthogonal input channels, the rank of the scattering tensor is always higher than that of the transmission matrix by a factor of $\frac{(N+1)}{2}$. Moon et al. showed that the higher rank of the scattering tensor offers significant advantages in numerous nonlinear optical operations, such as optical encryption, multichannel AND gates, and optical kernels in machine learning \cite{moon2023}. Their detailed work has been summarized in Ref. \cite{mujumdar2023}.

The advantage of using second-order nonlinear disordered media as a physical platform for large-scale optical computing was explicitly demonstrated by Wang et al.~\cite{wang2024}. They implemented a multilayer (deep) photonic neural network utilizing a disordered polycrystalline slab assembled from lithium niobate nanocrystals. A simplified representation of their experimental setup is shown in Fig.~\ref{fig:NL_computing}d. The disordered slab generated second-harmonic light at 400~nm when illuminated by pump femtosecond pulses at 800~nm. The slab provided strong multiple scattering at both the pump and SHG wavelengths, producing a linear and a nonlinear speckle pattern. Input information was encoded by spatial modulation of the pump phase profile. They showed that the use of the SHG speckle resulted in improved computing capabilities on various machine learning tasks, comprising image classification, univariate and multivariate regression, and graph classification. These striking results were explained by identifying the second-order optical nonlinearity as an extra (all-optical) activation function, in addition to the optoelectronic one implemented through intensity detection on the camera~\cite{saade2016random,rafayelyan2020prx}. In this description, the optical nonlinearity introduces an extra layer in the photonic neural network, improving the information storage capacity and the computing capabilities. 
A simplified representation of the two-layer optical neural networks obtained in this experiment is shown in Fig.~\ref{fig:NL_computing}f. The increased complexity of the network provides higher expressivity compared to the type of network implemented in linear disorder media shown in Fig.~\ref{fig:NL_computing}c.

\section{Conclusions and Future perspectives}
In summary, we have briefly reviewed recent developments in light transport and generation in disordered nonlinear media, emphasizing second-order nonlinearity. The emphasis is natural and arises from high coefficients of second-order nonlinearity that enable robust experimental measurements. 
The field of disordered systems, now termed linear disordered systems, has seen massive research activity in recent decades. Despite some enticing theoretical results, experimental endeavors into nonlinear systems have been lacking for a long time. The primary premise of this inactivity was the hyphenation of nonlinearity and phase-matching, referring to an unwritten diktat that absolute phase-matching was critical for any meaningful results on nonlinear optical materials. However, the growing realization of concepts such as random quasi-phase matching, controlled material fabrication and characterization, etc., led to a gentle ingress into the subfield of nonlinear disorder, which now holds the promise of transforming into a major, impaction research field with significant applications.

For ease of narration, we have classified the material space into three subdomains based on the scattering strength and nonlinear domain size. The system with the weakest structural disorder, namely, polycrystalline nonlinear media, remains transparent as the disorder is only due to random domain sizes and orientations. Various experimental results are described, and the diverse models for random quasi-phase matching are elaborately discussed. In these structures, the scattering is weak enough that the average direction of propagation of fundamental and nonlinear light does not change significantly, limiting the transport to a quasi-ballistic character.  Subsequently, we delve into the domain of multiple scattering nonlinear structures wherein the light diffuses throughout the sample in concurrence with nonlinear generation. A variety of experimental and theoretical results on simultaneous nonlinear generation, diffusion, and weak localization are described. A critical aspect of multiple scattering, namely speckle statistics, is described in the context of second-order nonlinear materials. Often, the creation of a single crystal of a nonlinear material with a sufficiently large size becomes a challenge, and multicrystalline structures or even powders need to be used. In such cases, new techniques for characterizing nonlinear coefficients have to be developed. In this review, we have described some of these techniques.

Rapid recent advances in controlled nanofabrication have led to the genesis of a designed nanodisorder, comprising assemblies of nonlinear nanoparticles involving both metals and dielectrics. In such instances, geometric structures often invoke Mie, whispering gallery, or plasmonic resonances to extract maximum nonlinearity in the assembly. Further, the existing technology of wavefront-shaping, extensively used in linear disorder, has also found significant usage in nonlinear disorder, where incident wavefronts of fundamental light have been modified to engineer the generation and propagation of nonlinear light. We enlist important aspects of this advance, including its applications, such as optical computing. 

Moving into the future, the emphasis will now change from obtaining efficient nonlinear signals from crystals to obtaining controllable nonlinear signals from structures. In the review, we have described that nanophotonic assemblies of $\chi^{(2)}$ materials are capable of generating controllable second harmonic signals. This could very well be the route towards a specialized 'horses-for-courses' approach in photonic technologies. Close-packed microsphere assemblies can enhance SHG with the help of Mie resonances, determined by their outer geometry. By engineering the outer geometry, numerous resonances can be achieved over a broad spectral range, potentially opening a new pathway to realizing random-quasi-phase-matching-based OPOs in miniaturized structures. Importantly, thanks to their small size, they could be easily integrated into any on-chip devices. This advancement can be a game-changer in quantum photonics. Specifically, many quantum photonic technologies rely on spontaneous parametric down-converted (SPDC) photons, essentially generated from bulk $\chi^{(2)}$-crystals. However, stringent phase-matching conditions restrict the generation of SPDC photons. Therefore, efficient and phase-matching free SPDC photon sources from engineered nanophotonic assemblies of $\chi^{(2)}$ crystals are expected to emerge in the coming years. Integrated into photonic chips, these microsources of biphotons can become a powerful platform towards quantum photonic devices.

On a different note, another domain of research that is currently experiencing explosive growth is artificial intelligence and deep neural networks. A critical feature in most neural networks is a filter that consists of a matrix of random elements. The strength of the network depends on the randomness and the number of possible random matrices involved in the development of the network. Disordered media have been shown to offer a physical platform for neural networks by creating speckle patterns that are essentially two-dimensional sets of totally random intensities. Physical implementations of neural network-based computation have been convincingly demonstrated using such disordered elements. Recently, the introduction of optical nonlinearity has been shown to enhance the computational capabilities of the network. On the one hand, the lower wavelengths generated through the disordered medium offer a richer speckle, and hence pack more `information' in the random matrix. On the other hand, the $\chi^{(2)}$-nonlinearity has been shown to act as an internal nonlinear activation function, speeding up loss elimination in the network. At the heart of this development has been the estimation of the 3D scattering tensor of the nonlinear disordered medium. It has been realized that these 3D tensors are much more vulnerable to the environment compared to linear 2D matrices. Overcoming this obstacle is a research area in itself. For example, transparent nonlinear glass samples doped with LiTaO$_3$ microcrystals have been shown to offer better stability \cite{cao2024} and may prove to be a solution to stable optical computing. Furthermore, the requirement of ultrashort pulsed lasers towards the generation of nonlinear signals is not energy-efficient. Recently, sum-frequency speckle from lithium niobate powder has been reported under CW illumination \cite{ni2023}. This promises a significant increase in energy efficiency in nonlinear optical computing processes. These studies are still young, and we realize that several potential contributions of a nonlinear disordered material in optical computing may exist and need to be researched.

\section*{Acknowledgments}
S.M. acknowledges the financial support from the Department of Atomic Energy, Government of India (12-R\&D-TFR-5.02-0200). 
R. Savo acknowledges support from the European Union - NextGenerationEU, project Comp-SECOONDO (MSCA$\_$0000079).
C. Conti acknowledges PRIN 2022597MBS PHERMIAC.

\bibliography{ref}

\end{document}